\documentclass[letterpaper]{nearbeam}
\usepackage{epsfig}
\begin{document}
   
\title{Coherent Nonlinear Phenomena in High Energy Synchrotrons:\\
 Observations and Theoretical Models}
\author{~P.~L.~Colestock, ~L.~K.~Spentzouris, and ~S.~I.~Tzenov\\
\it Fermi National Accelerator Laboratory \\
\it P.~O.~Box 500, Batavia, IL 60510, USA }
\newlength{\figwid}
\newlength{\figlen}
\maketitle
\medskip

\begin{abstract}
Nonlinear waves have been observed in synchrotrons for years but have received
little attention in the literature.  While pathological, these
phenomena are worth studying on at least two accounts.  First, the formation of 
solitary waves may lead to droplet formation that causes significant beam
halo to develop.  It is important to understand the conditions under which
such behavior may be expected in terms of the machine impedance.  Secondly, 
a variety of nonlinear processes are likely involved in the normal saturation
of unstable oscillations, leading to the possibility that low-level, but potentially
broadband fluctuation spectra may develop.  The resulting fluctuation spectra carry indirectly
the signature of the machine impedance.  In this work we review a number of observations
of nonlinear longitudinal waves made in Fermilab accelerators, 
and make a first attempt to develop
appropriate theoretical models to explain these observations.  
\end{abstract}

\par
\section {Introduction}
\par
Over the years, nonlinear wave phenomena have received scant attention
in high energy synchrotrons, in part, because of the mathematical 
difficulty of this subject, but also due to the fact that nonlinear 
wave motion is usually associated with a pathological state of an
accelerator that is best to be avoided.  While this is indeed true for 
the most violent nonlinear effects, a broad class of low-level processes
may be playing a role in many present machines, and the drive for ever higher
beam intensities may lead to the widespread occurrence of nonlinear 
wave phenomena.
\par
In particular, in the case of the dynamical behavior in the 
vicinity of an intense,
stored beam, we are interested in the formation of beam halo,
either as a diffuse cloud, represented by a departure from a Gaussian
distribution, or as droplets which may occur in a type of phase transition
at beam's edge due to coherent modes.  In addition, it is useful to study
the formation of an equilibrium state, if it exists, between a broad spectrum
of marginally stable modes and some weak dissipative mechanisms that can lead
to a saturated state of low-level turbulence, which, in turn, can affect the
rate at which the halo population is generated.  These phenomena can be
expected to be most prevalent in hadron rings owing to the weakness of
the damping mechanisms, and our attention in this paper is focussed on
this case.
\par
While these subjects are mathematically complex, a rich literature already
exists in the field of plasma physics that deals with these questions,
although the interparticle force predominantly considered in this
literature is due to space charge alone.  At the relativistic 
energies typical in modern accelerators, the interaction
between particles is dominated by wall image currents, i.e. the wakefields,
which complicates the nature of the interaction, but can also lead to a 
wider variety of wave phenomena. It is our aim in this work to highlight 
observations of nonlinear wave phenomena in high-energy synchrotrons and
to point out methods of analysis from plasma physics that can be applied
to the study of these topics.
\par
The types of wave behavior in beams may be classified according to the
degree of nonlinearity, in parallel with the concepts in plasma physics.
In the linear regime, a resonant mode can be driven resulting in a
response at the drive frequency which is characteristic of the beam
intensity and the nature of the wakefield, or impedance, of the machine.
When detected by a suitable pick-up, the driven response can shed
light on the properties of the wakefields and the proximity of the 
beam to the stability threshold.  This socalled transfer function
method \cite{hofmann} is widely used to study accelerator stability.
\par
If an accelerator is operated just 
above its stability limit, the most unstable mode is driven into 
exponential growth by the wakes, reaching a saturated, though marginally 
stable, state as the beam distribution is altered by the growing
waves.  If the spectrum of unstable modes is sufficiently broad, the
phase of the perturbation is effectively random, and the interaction of
waves and particles leads to particle diffusion in phase space, known
in plasma physics as quasi-linear diffusion \cite{kennel}.  The analog
in beams, known as the 'overshoot' phenomenon, has been studied
\cite{chin},\cite{bogacz}, although the applicability of this model
is unclear owing to the typically narrow unstable spectrum found in
many storage rings.  A recent numerical study of this phenomenon
that shows the complexity of the interaction is found in ref. 
\cite{gerasimov}.
\par
However, particularly in hadron rings where the absence of synchrotron
damping allows virtually unimpeded mode growth, unstable 
waves can grow to finite
amplitudes that permit a significant fraction of the beam to become
trapped in its own wake.  The resulting wave motion couples to the 
trapped particles in such a way as to give rise to slowly damped
oscillations.  This phenomenon is known as nonlinear Landau damping
in the plasma physics literature \cite{oneil1}
(in comparison to linear Landau damping which is part of the linear
beam response).  It is to be expected that where a discrete spectrum
of unstable waves can occur, as is often the case in a synchrotron,
that nonlinear Landau damping can play an important role.
\par
At the next higher level of nonlinear interaction, 
coherent modes can resonantly interact in a process known as the three-
wave interaction \cite{nishikawa}.  This leads typically to a cascade
in frequency which, due to the harmonic character of many modes in
storage rings, readily occurs and can cause a broadening 
of the original unstable
spectrum.  This phenomenon has been studied in the simple case of
longitudinal oscillations in a coasting beam, \cite{spentzouris}
and it can be expected that
similar wave-wave coupling can occur in the transverse plane and in
bunched beams as well, albeit with different resonance conditions.
\par
If the coherent motion is particularly violent, and sufficiently
dissipative, then the trapped portion of the beam can self-extract from
the core of the beam distribution, forming droplets at the beam edge that
can be self-sustaining.  These are, presumably, a form of solitary wave,
or soliton, which is perhaps unique to a high-energy synchrotron due 
to the complex character of the wake field.  Such solitary waves may be
a primary producer of halo particles for weakly-damped hadron rings.
\par
In general, we are interested in the final state of these various
nonlinear interactions: the condition where the coherent modes reach
marginal stability through either a change in the beam distribution,
or through frequency spreading of the spectral distribution.  In the
latter case, the phonons themselves can be thought of as comprising a 
fluid which comes into equilibrium, the details of which depend on 
the inter-phonon interaction.  In the plasma physics literature,
scaling laws for the resulting turbulent fluctuation spectrum have
been derived (\cite {zakharov} and references contained therein).  For
our purposes, we would like to understand how aspects of the machine
impedance, and therefore the detailed design, contribute to the 
form of the equilibrium turbulent spectrum, if it exists.
\par
In this paper we review the observations made at Fermilab
\cite{spentzouris} in stored high energy hadron beams and 
compare the observations with numerical simulations.  
Our experimental studies, and thus our theoretical work, have
been focussed on the phenomena in perhaps the simplest of all cases, that 
of longitudinal oscillations of a coasting, or unbunched, beam in a 
storage ring.  As such, the surface
of this subject has only been scratched, and our aim here is to outline 
the steps that would have to be taken to study any of the many other
possible situations where nonlinear waves can occur.  Moreover, we would
like to underscore the importance of understanding turbulence in beams
that we feel will be playing an increasingly important role as beams are
more commonly run close to, or even above, their linear stability boundaries.
\par
\section {Review of Basic Phenomena}
\subsection{Stability in Particle Beams}
In the case of a high-energy stored beam, the growth of coherent wave
motion is normally undesireable.  Wakefields can drive such waves, though
the mode growth is counteracted by damping due to the spread in frequencies
of the individual particles making up the beam, and this damping effect 
was first derived for a plasma, known as Landau damping \cite {stix}. A
well known technique for determining the linear stability boundary of a 
beam is to excite driven oscillations on the beam and to monitor the amplitude
and phase of the beam's response, which includes the effects of wakefields.
This technique, known as a beam transfer function, \cite{hofmann}, yields
for longitudinal motion in a coasting beam a response of the form 
\begin{equation}
R(\Omega) =  \frac{1}{\frac{i(e \omega_s)^2}{2 \pi} \int_{-\infty}^{\infty} 
\frac{\frac{\partial f_o}{\partial \epsilon}}
{\Omega - m (\omega_s + k_o \epsilon)} d \epsilon} + Z(\Omega) 
\label{eq:response}
\end{equation}
where $ \omega_s $ is the harmonic revolution frequency, $f_o$ is the
longitudinal particle distribution function, $\varepsilon$ is the energy
deviation, $k_o$ is the frequency dispersion factor and $Z(\Omega)$ 
is the machine impedance.  
This function is directly related to the dispersion
relation for longitudinal modes given by 
\begin{equation}
D_m(\Omega) = 1 + Z(\Omega) \frac{i(e \omega_s)^2}{ 2 \pi}
 \int_{-\infty}^{\infty} \frac{\frac{\partial f_o}{\partial \epsilon}}
{\Omega - m (\omega_s + k_o \epsilon)} d \epsilon 
\label{eq:dispersion}
\end{equation}
\begin{figure}[hbtp]
\centerline{\psfig{file=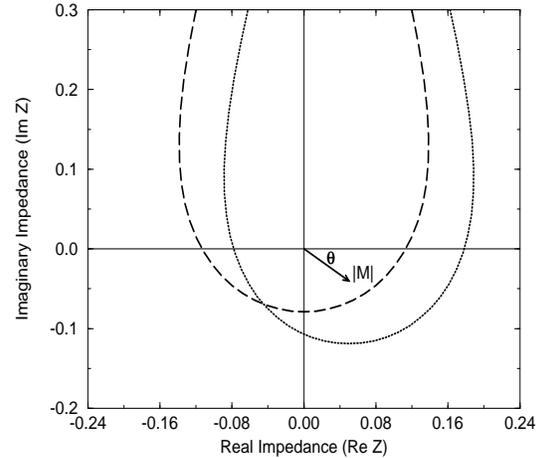,height=2.75in,width=3.0in}}
\caption
[Theoretical shift of the beam response due to an impedance.]
{Theoretical shift of the beam response due to an impedance.  
The curve centered on the origin (dashed) is the response 
when there is no impedance, and the displaced curve (dotted) is the 
response when there is an impedance of \mbox{$(Z_{x},Z_{y})=(.05,-.04)$}.  
The magnitude of the impedance is 
\mbox{$|{\rm M}|=\sqrt{Z_{x}^{2}+Z_{y}^{2}}=.064\Omega$}, 
and the phase is \mbox{$\theta=-39^{\circ}$}.  The beam distribution used 
was Gaussian in energy, and the beam parameters were arbitrary.}
\label{figure:tfunc}
\end{figure}
\noindent

The stability boundary can be depicted in the impedance plane as the
curve for which $Im( \Omega ) = 0$, as shown in Fig.\ref{figure:tfunc}.
The machine impedance can be extracted from the measurements as an offset 
of the centroid of the stability curve, provided the beam distribution
is known, assumed to be Gaussian here.

\subsection{The Three-Wave Interaction}
Weakly nonlinear processes are described using 
the same techniques as in linear stability theory, with the exception that 
a second-order frequency mixing term is included in the description of the 
dynamics.  The effect of the frequency-mixing leads to a resonant coupling
phenomenon by which modes at two separate frequencies couple to produce a
response at a third frequency, a process known as three-wave or parametric
coupling.  The process is characterized by selection rules such that

\begin {equation}
\omega_1 = \omega_2 + \omega_3 
\label{equation:selection}
\end{equation}
corresponding to conservation of energy among the waves.  A similar
condition applies to the mode wavenumbers, corresponding to 
conservation of momentum.  Due to the periodicity in a ring, 
this condition can be readily satisfied for a large number of normal modes.
We have studied the coupling for longitudinal modes theoretically 
and found three-wave coupling obeys a dispersion relation that couples
the linear response of harmonic m and m-n through an idler mode at
harmonic n.
\begin{eqnarray}
\lefteqn{
D_{m}(\Omega)D_{m-n}(\Omega-\Omega_{0})  = }
\nonumber 
\\
& &
\frac
{ I_{0}^{2}Z_{m}(\Omega)Z_{m-n}(\Omega-\Omega_{0})V_{0}V_{0}^{*}\beta^{8} }
{ 64\pi^{5}m^{2}(m-n)^{2}\eta^{4}
\left( \frac{\sigma_{\varepsilon}}{E_{0}}\right)^{8}(E_{0}[eV])^{4} } 
\nonumber \\
& \times &
\int_{-\infty}^{\infty}\frac{xe^{-x^{2}}}
{[\varepsilon-\xi_{1}]
[\varepsilon-\xi_{2}]^{2}}\,dx
\nonumber \\
& \times &
\int_{-\infty}^{\infty}\frac{xe^{-x^{2}}}
{[\varepsilon-\xi_{1}]^{2}
[\varepsilon-\xi_{2}]}\,dx
\label{equation:modedisp}
\end{eqnarray}
where the drive frequency is at $\Omega_o = n \omega_s$ and

\begin{eqnarray*}
x & = & \frac{\varepsilon}{\sqrt{2}\sigma_{\varepsilon}} ,\\ 
\xi_{1} & = & \frac{1}{\sqrt{2}\sigma_{\varepsilon}mk_{0}}(\Omega-m\omega_s) ,\\
\xi_{2} & = & \frac{1}{\sqrt{2}\sigma_{\varepsilon}(m-n)k_{0}}(\Omega
-\Omega_{0}-(m-n)\omega_s) , 
\end{eqnarray*}
\noindent
and $V_o$ is the drive amplitude, $I_o$ is the beam current, $\eta$
is the slip factor, $\frac{\sigma_\epsilon}{E_o}$ is the fractional
energy spread and $E_o$ is the beam energy.

\begin{figure}[t]
\setlength{\figlen}{4.75in}
\setlength{\figwid}{3.0in}
\centerline{\psfig{file=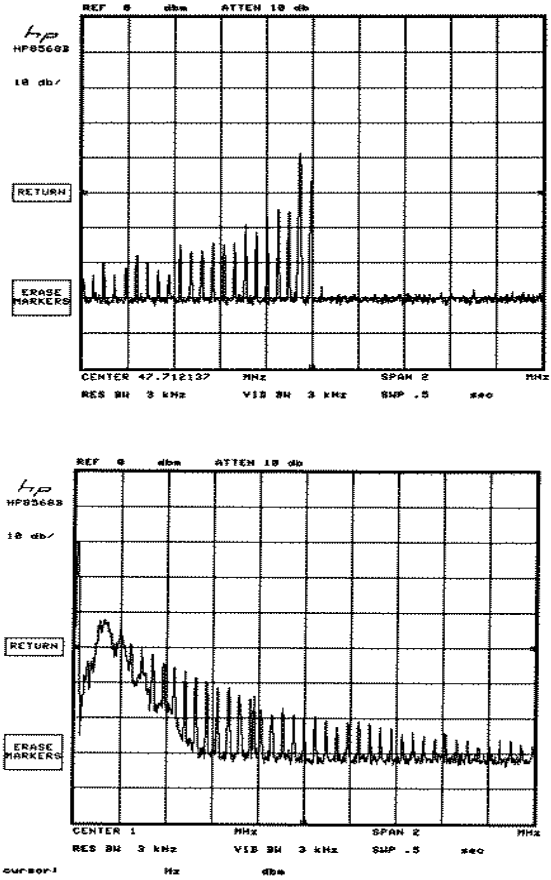,height=\figlen,width=\figwid}}
\caption{
Three-wave coupling spectrum for longitudinal modes in a coasting
beam in the Tevatron, 150 GeV beam.  Excitation at h = 1000, (47.712 MHz)
as shown in the upper graph, led to successive excitation 
of lower sidebands accompanied by low frequency modes, shown 
in the lower graph, which satisfy the selection rule.  The lower graph
begins at zero frequency and in each figure the frequency span is 2 MHz.
The vertical amplitudes are in arbitrary units but the scales are logrithmic.
}
\label{figure:spectrum}
\end{figure}

The implication of Eq. \ref{equation:modedisp} is that three-wave coupling
is most likely near the stability threshold for any of the modes involved.
The selection rule Eq. \ref{equation:selection} leads to a
single-sided coupling, which was observed experimentally, as shown in Fig.
\ref{figure:spectrum}.

An interesting issue to investigate is how the power in the excited
modes varies in time, especially in the presence of damping.  Experimental
observations indicate that a very regular cascade toward lower
frequencies takes place, evidently due to successive three-wave 
coupling events.  This behavior is typical for a dissipative system
with sufficiently high mode density, and may be described by the
following system of equations for the mode amplitudes.

\begin{equation}
\frac{\partial A_{m} }{\partial t} = 
s_{m}e^{i\phi}V_{mnk}A_{n}A_{k}
\label{eq:adot}
\end{equation}
where the matrix element of the interaction has been symbolized as $V_{mnk}$, 
and is defined by as the following,
\newpage
\begin{eqnarray}
\lefteqn{
V_{mnk}   = } \nonumber \\ 
& &
\!\!\!\!\!\!\!\!\!\!\!\!
\frac
{
-i\frac{2(e\omega_s)^{3}}{\sqrt{\epsilon_{0}}(2\pi)^{2}} Z
\int_{-\infty}^{\infty}\frac{\partial}{\partial \varepsilon}
\! \left(
\frac{\frac{\partial f_{0}}{\partial \varepsilon}}
{\omega_{k}-(k)\omega(\varepsilon)}
\right)
\frac{1}{[\omega_{m}-m\omega(\varepsilon)]}
\,d\varepsilon
}
{
\left[
\left|\frac{\partial D_{m}}{\partial \omega_{m}}\right|
\left|\frac{\partial D_{n}}{\partial \omega_{n}}\right|
\left|\frac{\partial D_{k}}{\partial \omega_{k}}\right|
\right]^{\frac{1}{2}}
}
\label{equation:amplitude}
\end{eqnarray}

It is worthwhile to note that as the multiplicity of modes becomes 
sufficiently dense, the coupling between waves governed by Eq. 
\ref{equation:amplitude} can lead to a solitary wave phenomenon
\cite{fermi}, and this subject will be described further in a later 
section.  In the above mentioned work, only the interaction of 
longitudinal modes has been considered.  It is also reasonable to
expect that transverse modes can be coupled, especially where
nonlinearities can play an important role, such as in the beam-beam
interaction.  This should be a fruitful area for further study.

\subsection{Nonlinear Landau Damping}

Sufficiently large wakefields disturb an initially smooth particle 
distribution by trapping particles within the potential wells of the
waves generated.  The particle motion decoheres with a time constant
that is significantly longer than the inverse frequency spread, or
linear Landau damping time.  The trapped particles undergo synchrotron
oscillations in the self-generated potential wells, alternately exchanging
incremental energy with the wakefields.  The combination of energy dispersion
of the particles and the nonlinearity of the voltage waveform eventually
causes phase mixing of the coherent motion.  This nonlinear damping process
is called nonlinear Landau damping, and was first studied in plasma physics.
\cite{oneil1}, \cite{dawson} -  \cite{oneil2}.
\par
 
\begin{figure}[t]
\setlength{\figlen}{2.5in}
\setlength{\figwid}{3.0in}
\centerline{\psfig{file=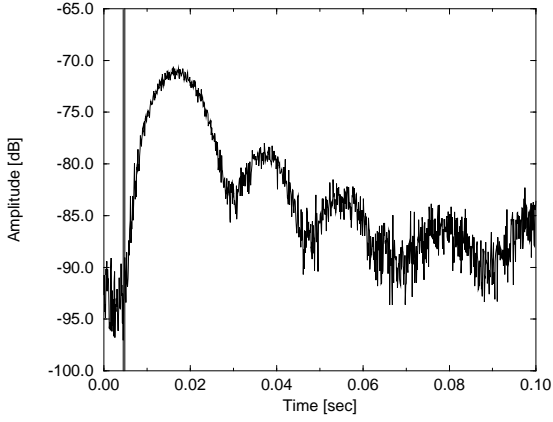,height=\figlen,width=\figwid}}
\caption{Power versus time at h = 105 in response to a 0.5 msec drive
pulse.  The placement and duration of the drive pulse has been 
drawn in for reference.  An impulse excitation leads to slowly
decaying amplitude oscillations as trapped particles exchange energy
with the wakefield.}
\label{figure:bounce}
\end{figure}

Experiments were carried out in the Fermilab Main Ring which clearly
showed the signature of nonlinear Landau damping.  In these studies, a short
pulse of rf power was applied to the beam using an rf cavity at h=106 
(5.03 MHz).  The resulting response showed a characteristic response whose
envelope decayed not exponentially but in an oscillatory manner, as shown
in \ref{figure:bounce}.  This behavior is attributed to the exchange of
energy between trapped particles and waves as described above.
Both analytic \cite{oneil2}, \cite{dawson}, \cite{canosa}, and 
numerical work \cite{oei}, on nonlinear Landau damping has been 
carried out which descibes the behavior we have observed.  This
result will also be discussed further in a later section on simulations.
\par
It has also been pointed out \cite{oneil2} that the advent of particle 
bunching is accompanied by the appearance of coherent power in higher 
harmonics of the fundamental frequency of the wakefield as the trapped
particle bunches compress within the potential wells.  Such behavior has
indeed been observed in the experiments described above \cite{spentzouris}.
This compression of the bunch length is essentially wavefront 
steepening which is a prerequisite for the formation of solitons.

\subsection{ Solitary Waves.}

The formation of solitons in a beam is of interest, since solitons may well be the
vehicle which carries coherent energy in a highly turbulent state that might
occur in a beam with weak damping.  A vast literature on solitons
in various media exists \cite{drazin} - \cite{davidson}, though little effort has been 
given to this subject in ultra-relativistic beams.  In particular, solitons
in plasmas have been studied extensively, \cite{sagdeev}, \cite{davidson},
which depend on the particular nonlinearity introduced by the Coulomb
force, i.e. space charge.  A similar space charge limit was studied for
a coasting beam \cite{bisognano}, \cite{fedele}, and for a resonator impedance,
\cite{arci2}, leading to the possibility that solitons may exist in
high energy beams under certain conditions.
\par
Since a wakefield force is fundamentally more complex than the space
charge force, it can be assumed that the characteristics of a soliton
will also be unique to the case of a high-energy beam.  In particular,
it is interesting to know what the impact of the wakefield dissipation 
has on the solitary wave behavior.  Results from a variety of beam
experiments suggest that long-lived solitary structures may form and
extract themselves from the core of the beam over long times \cite{cosmo},
\cite{carrigan}.  In other work, transient solitary waves seem to
appear \cite{spentzouris}.  In all cases, solitons may be viewed as
phase-space droplets  that appear in the beam, and under some conditions,
give rise to a phase-transition and clumpy halo formation.  From this point
of view, it is valuable to understand their dynamics.
\par
To this end we sketch here the results of an analytic study of 
longitudinal solitons on a coasting beam due to a general resonator impedance.
This reperesents the simplest possible scenario
for understanding such phenomena, will illustrate the mathematical 
procedure and serves as the starting point for more complex situations.
For the reader's sake we note that many steps have been omitted from the
following derivation for reasons of space.  Full details will be given
in forthcoming work \cite{tzenov}.
The model equations for the dynamics are given by the following system

\[
\frac{\partial f}{\partial T}+v\frac{\partial f}{\partial \theta }+\lambda
_1V\frac{\partial f}{\partial v}=0,
\]

\begin{equation}
\frac{\partial ^2V}{\partial T^2}+2\gamma \frac{\partial V}{\partial T}%
+\omega ^2V=\frac{\partial I}{\partial T},
\end{equation}

\[
I\left( \theta ;T\right) =\int dvvf\left( \theta ,v;T\right) , 
\]

\noindent where $f\left( \theta ,v;T\right) $ is the longitudinal
distribution function, $V$ is the voltage on a resonator of $\gamma =\frac
{\omega}{2Q}$, $\omega =\frac{\omega _r}{\omega _s}$ $\omega _r$
being the resonator frequency, and $I$ is the instantaneous beam
current. Time $t$ has been normalized as $T=\omega _st$. Furthermore $%
v=\frac 1{\omega _s}\frac{d\theta }{dt}=1+\frac{k_o\varepsilon }{\omega _s}$
is the dimensionless angular velocity of a beam particle and

\[
\lambda _1=\frac{e^2Rk_o\gamma }\pi ,
\]

\noindent where $R$ is the resonator shunt impedance. Using standard moment
techniques \cite{klimon} on the above equations, we may pass over to
the hydrodynamic picture of longitudinal beam motion and start from the
system of gas-dynamic equations

\[
\frac{\partial \rho }{\partial T}+\frac \partial {\partial \theta }\left(
\rho u\right) =0, 
\]

\begin{equation}
\frac{\partial u}{\partial T}+u\frac{\partial u}{\partial \theta }=\lambda V-%
\frac{\sigma _v^2}\rho \frac{\partial \rho }{\partial \theta },
\end{equation}

\[
\frac{\partial ^2V}{\partial T^2}+2\gamma \frac{\partial V}{\partial T}%
+\omega ^2V=\frac \partial {\partial T}\left( \rho u\right) ,
\]

\noindent where $\rho $ and $u$ are the density and the mean velocity
moments of the distribution, respectively, (the variables $\rho $ and $V$
have been appropriately scaled) and $%
\lambda =\rho _o\lambda _1$.   ($\rho _o$ = constant is the equilibrium
beam density.)  Using a renormalization group approach \cite{tzenov}, 
\cite{goldenfeld}, we may derive a set of amplitude equations for the
rescaled beam density $R_o$, the current velocity $u_o$ and the mode envelope
function $E$.

Before proceeding, we would like to examine the stability problem of
stationary waves in this system, which can be done without a formal solution
of the amplitude equations. The approach, introduced by Sagdeev \cite
{sagdeev}, is to look for forms of the nonlinear equations which correspond
to harmonic motion in an effective potential well. Such states, if they
exist, are conjectured to be allowed solitary waves in the nonlinear system.
To this end let us write down the full system of amplitude equations, which
after appropriate scaling reads as

\[
\frac{\partial \widetilde{\rho }}{\partial \tau }+\frac \partial {\partial
\Theta }\left( \widetilde{\rho }\widetilde{v}\right) =0, 
\]

\begin{equation}
\frac{\partial \widetilde{v}}{\partial \tau }+\widetilde{v}\frac{\partial 
\widetilde{v}}{\partial \Theta }=-c_u^2\left( \frac 1{\widetilde{\rho }}%
\frac{\partial \widetilde{\rho }}{\partial \Theta }+\frac{\partial \left|
\psi \right| ^2}{\partial \Theta }\right) ,
\end{equation}

\[
i\frac{\partial \psi }{\partial \tau }+i\gamma b\psi +\left( 1-\widetilde{%
\rho }\right) \psi =-\left( 1-\frac{2i\gamma }{\omega _o}\right) \frac{%
\partial ^2\psi }{\partial \Theta ^2}+ 
\]

\[
+\frac{2i}{\omega _ob}\left( 1-\frac{i\gamma }{\omega _o}\right) \frac{%
\partial \left( \widetilde{\rho }\widetilde{v}\psi \right) }{\partial \Theta 
}+\frac i{\lambda b^2}\psi \frac{\partial \widetilde{\rho }}{\partial \tau }%
, 
\]

\noindent where

\[
\widetilde{\rho }=1+R_o\qquad ;\qquad \widetilde{v}=\frac{ab\omega _o}%
2\left( 1+u_o\right) ,
\]

\[
\psi =\frac{\left| \lambda \right| }{\omega _o\sigma _v}Ee^{-\gamma T}
\]

\noindent are the new (rescaled) dependent variables. The coefficients
entering the above expressions are specified as follows:

\[
a=\frac 2{\sqrt{\sigma _v^2+3}}\qquad ;\qquad b=\frac{2\omega _q}\lambda ,
\]

\[
c_u=\frac{a\omega _o\omega _q\sigma _v}{\left| \lambda \right| },
\]

\noindent where $\sigma _v$ is the normalized beam energy spread. The new
independent variables (time $\tau $ and azimuthal position $\Theta $) are
given by

\[
\tau =\frac Tb\qquad ;\qquad \Theta =\frac{a\omega _o\theta }2.
\]

\noindent Moreover in the above set of equations the following notations
have been adopted

\[
\omega _o^2=\omega ^2-\lambda \qquad ;\qquad \omega _q^2=\omega _o^2-\gamma
^2,
\]

\noindent In order to proceed, we have to further assume that the resonator
is weakly damped, namely the high-$Q$ case. For this case $\left( \gamma
=0\right) $ the ansatz

\[
\widetilde{\rho }=\widetilde{\rho }\left( z\right) \qquad ;\qquad \widetilde{%
v}=\widetilde{v}\left( z\right) , 
\]

\[
\psi =A\left( z\right) e^{i\left[ \left( a+v_o\right) z/2+\Omega \tau
\right] }\quad ;\quad z=\Theta -v_o\tau 
\]

\noindent leads to a system of differential equations for $\widetilde{\rho }$%
, $\widetilde{v}$ and $\psi $ admitting the following integrals of motion

\[
C_1=\widetilde{\rho }\left( \widetilde{v}-v_o\right) , 
\]

\begin{equation}
C_2=\frac{C_1^2}{2\widetilde{\rho }^2}+c_u^2\left( A^2+\ln \widetilde{\rho }%
\right) ,
\end{equation}

\[
2E=\left( \frac{dA}{dz}\right) ^2+\left[ 1-\Omega +\frac 14\left(
a+v_o\right) ^2\right] A^2- 
\]

\[
-1+\widetilde{\rho }+\frac{C_1^2}{c_u^2\widetilde{\rho }}. 
\]

\begin{figure}[t]
\setlength{\figlen}{2.25in}
\setlength{\figwid}{3.0in}
\centerline{\psfig{file=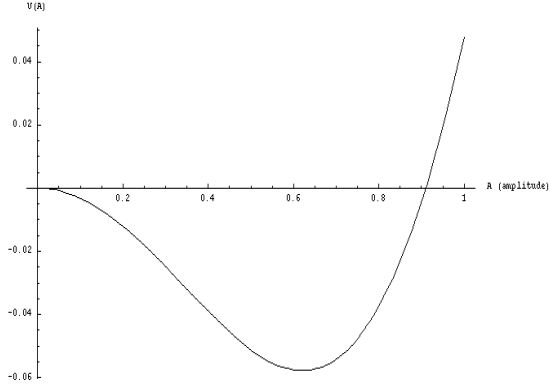,height=\figlen,width=\figwid}}
\caption{Pseudo-potential based on the nonlinear wave
equations for a coasting beam.  A minimum in this potential indicates
the possibility for solitary waves (cavitons) to form.}
\label{figure:potential}
\end{figure}

\noindent These integrals of motion suggest that the stability of stationary
waves can be equivalently described in terms of motion of a single particle
in a (pseudo)- potential well. Indeed, the function

\[
U\left( A\right) =\left[ 1-\Omega +\frac 14\left( a+v_o\right) ^2\right]
A^2- 
\]

\begin{equation}
-1+\widetilde{\rho }+\frac{C_1^2}{c_u^2\widetilde{\rho }}
\end{equation}

\noindent provided $\widetilde{\rho }$ is expressed in terms of $A$ from the
second integral, comprises a pseudo-potential function. In Fig. \ref{figure:potential}
we show $U\left( A\right) $ for the simplest case of constant current velocity $%
\widetilde{v}=v_o$ $\left( C_1=0\right) $. We note that a minimum in this
pseudo-potential corresponds to solitary waves that can be effectively
``trapped'' in this potential well. 

In the following, we can proceed to find
approximate closed-form solutions of these nonlinear equations, which
will allow us to explicitly find the time behavior of the solitary waves in the
presence of dissipation.
Eliminating of the current velocity from the complete set of amplitude
equations and expressing $\widetilde{\rho }$ in terms of $\left| \psi
\right| ^2$

\begin{equation}
\widetilde{\rho }=1-\left| \psi \right| ^2
\end{equation}

\noindent we finally arrive at the damped nonlinear Schr\"odinger equation

\[
i\frac{\partial \psi }{\partial \tau }+i\gamma b\psi =
\]

\begin{equation}
-\left( 1-\frac{2i\gamma }{\omega _o}\right) \frac{\partial ^2\psi }{%
\partial x^2}+\frac{a\gamma }{\omega _o}\frac{\partial \psi }{\partial x}%
-\left| \psi \right| ^2\psi ,
\label{equation:soliton}
\end{equation}

\noindent where

\[
x=\frac a2\left( \omega _o\theta +\frac{2T}b\right) .
\]

\begin{figure}[t]
\setlength{\figlen}{2.75in}
\setlength{\figwid}{3.0in}
\centerline{\psfig{file=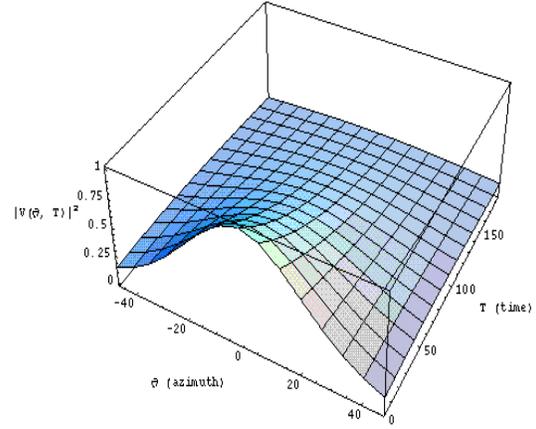,height=\figlen,width=\figwid}}
\caption{Voltage amplitude evolution of the solitary wave due to
a resonator impedance.  In the frame of the wave, the amplitude persists
for long times but eventually damps as the soliton decelerates away
from the beam core, and hence, the resonator's resonant frequency.}
\label{figure:amplitude}
\end{figure}

\begin{figure}[t]
\setlength{\figlen}{2.75in}
\setlength{\figwid}{3.0in}
\centerline{\psfig{file=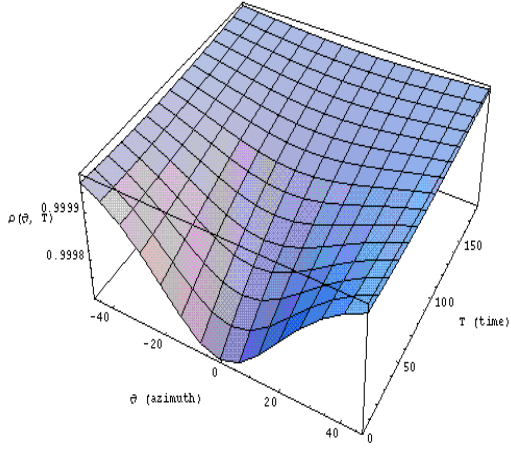,height=\figlen,width=\figwid}}
\caption{Caviton density evolution.  A density depression associated
with the solitary wave decreases in amlitude at long times as the
wave amplitude itself decreases.}
\label{figure:caviton}
\end{figure}

\noindent
Eq. \ref{equation:soliton} admits closed form solutions that indicate
solitary waves can exist, but due to the dissipation in the model,
eventually disappear after initial generation.  We interpret this
behavior as a gradual shrinking of the potential well that occurs
when the trapped particles have decelerated sufficiently from the
the resonantor frequency.  
The results for the voltage amplitude and soliton (caviton) density
are shown in Figs. \ref{figure:amplitude} and \ref{figure:caviton}
respectively.  We note that a similar equation has been
derived \cite{fedele} from an entirely different perspective.
It is also worth noting that Eq. \ref{equation:soliton} is a
special case of the complex, cubic Ginzburg-Landau equation
\cite{cross}, widely used to study various pattern formation
phenomena and coherent structures. 

\subsection{ Turbulence.}

The study of turbulence in beams is valuable primarily because it may
be a universal phenomenon, at least at low levels, which plays a role
in determining the limiting phase-space density in any machine.  The
effect has likely been small in machines well below their stability
thresholds, however, as intensities have been pushed closer to stability
limits, nonlinear wave interactions can occur which lead to a marginally
stable equilibrium.  A first attempt at determining the fluctuation
spectrum for a beam was obtained by considering the equilibrium state
of a Gaussian beam \cite{parkhom}.  The resulting spectrum was related
to the linear dielectric function and showed that the fluctuation
density would be strongly peaked for cases near the linear stability
limit.  It is our conjecture that such a situation may have
occurred in the Fermilab Tevatron during recent attempts to realize
stochastic cooling of bunched beams \cite{jackson}.  A broad, stationary
spectrum of fluctuations was observed at many times the expected Schottky,
or shot noise, levels.  The harmonic generation observed in the Fermilab 
Main Ring mentioned above is consistent with these observations as well.  
It is our aim in this paper to outline a theoretical approach to
understand the formation of an equilibrium spectrum and to understand 
the spectral amplitude dependence on the character of the machine
impedance.
\par
The approach taken is to develop a statistical description of fluctuations
for an ensemble of coupled modes.  This may be viewed as a development
of the amplitude equations associated with coupled modes, as in Eq.
\ref{equation:amplitude}.  The interaction between modes may be three-wave, 
as given in \ref{equation:amplitude}, or higher-order.  In this
work, we outline a general procedure, but keep only interactions up to
the three-wave level.  The result will be a scaling law for the envelope
of the fluctuation spectrum.
\par
The starting point for our analysis is the system of equations for the
fluctuation $\delta N$ of the microscopic phase space density and the
fluctuation $\delta V$ of the voltage \cite{zakharov}
\begin{eqnarray*}
\lefteqn{
\left( \frac \partial {\partial \theta }+v\frac \partial {\partial \sigma
}+\lambda V\frac \partial {\partial v}\right) \delta N = } \nonumber \\
& &
-\lambda n\frac{%
\partial f}{\partial v}\delta V-\lambda \frac \partial {\partial v}\left[
\delta N\delta V-\left\langle \delta N\delta V\right\rangle \right] , 
\end{eqnarray*}

\[
\frac{\partial ^2\delta V}{\partial \sigma ^2}-2\gamma \frac{\partial \delta
V}{\partial \sigma }+\omega ^2\delta V=-\frac{\partial \delta I}{\partial
\sigma }, 
\]

\[
\delta I=\int dv\left( 1+v\right) \delta N\left( \sigma ,v;\theta \right) 
\]

\noindent written in the variables $\sigma =\theta -\omega _st$ and $%
v=k_o\varepsilon /\omega _s$. Fourier transforming the above equations and
using the concept of slowly varying amplitude of weakly nonlinear waves one
obtains the following equation

\[
\left( \frac \partial {\partial \theta }+\Omega _g\frac \partial {\partial
\sigma }-\Gamma _k\right) \delta V_k=-\frac i{\left( \frac{\partial \epsilon 
}{\partial \Omega }\right) _{\Omega _k}} 
\]

\[
\times \sum_{k_1+k_2=k}\kappa _2\left( k,\Omega _{k_1}+\Omega_{k_2};k_2,\Omega _{k_2}\right) 
\]

\begin{equation}
\times \left( \delta V_{k_1}\delta
V_{k_2}-\left\langle \delta V_{k_1}\delta V_{k_2}\right\rangle \right)
e^{i\theta \left( \Omega _k-\Omega _{k_1}-\Omega _{k_2}\right) }+... 
\end{equation}

\noindent for the amplitudes $\delta V_k$, where

\[
\epsilon \left( k,\Omega \right) =1+i\lambda nZ\left( k\right) \int \frac{%
dv\left( 1+v\right) }{\Omega -kv+io}\frac{\partial f}{\partial v}, 
\]

\noindent is the dielectric permittivity and

\[
\kappa _2\left( k,\Omega ;k_1,\Omega _1\right) = 
\frac{n\lambda ^2Z\left(k\right) }{2\gamma }\int dv\left( 1+v\right) 
\]
\[ 
\times\frac 1{\Omega -kv+io}\frac
\partial {\partial v}\left( \frac 1{\Omega _1-k_1v+io}\frac{\partial f}{%
\partial v}\right) 
\]

\noindent is the second order susceptibility of the beam. Here

\[
Z\left( k\right) =\frac{ik}{k^2-\omega ^2+2i\gamma k} 
\]

\noindent is the familiar resonator impedance function and

\[
\Omega _g=-\left\{ \frac{\partial Re\left[ \epsilon \left( k,\Omega \right)
\right] }{\partial k}\left[ \frac{\partial Re\left[ \epsilon \left( k,\Omega
\right) \right] }{\partial \Omega }\right] ^{-1}\right\} _{\Omega =\Omega
_k} 
\]

\[
\Gamma _k=\left\{ Im\left[ \epsilon \left( k,\Omega \right) \right] \left[ 
\frac{\partial Re\left[ \epsilon \left( k,\Omega \right) \right] }{\partial
\Omega }\right] ^{-1}\right\} _{\Omega =\Omega _k} 
\]

\noindent are the group velocity of waves and the damping factor
respectively. For the slowly evolving part $A_k$ of the wave amplitude $%
\delta V_k$ it is straightforward to derive the equation

\[
\left( \frac \partial {\partial \theta }+\Omega _g\frac \partial {\partial
\sigma }+i\omega _k-\Gamma _k\right) A_k= 
\]

\[
=i\sum_{k+k_1=k_2+k_3}S\left( k,k_1,k_2,k_3\right) 
\]

\begin{equation}
\times \left(
 A_{k_1}^{*}A_{k_2}A_{k_3}-A_{k_2}\left\langle
A_{k_1}^{*}A_{k_3}\right\rangle -\left\langle
A_{k_1}^{*}A_{k_2}A_{k_3}\right\rangle \right) ,  
\label{slwvamp}
\end{equation}

\noindent where

\[
S\left( k,k_1,k_2,k_3\right) =\frac{\kappa _2\left( k-k_2,\Omega
_{k_3}-\Omega _{k_1};k_3,\Omega _{k_3}\right) }{\sqrt{\omega ^2-\lambda n}%
\left( \frac{\partial \epsilon }{\partial \Omega }\right) _{\Omega _k}\left( 
\frac{\partial \epsilon }{\partial \Omega }\right) _{\Omega _{k-k_2}}} 
\]
%
\[
\times [ \kappa _2 ( k,\Omega _{k_2}+\Omega _{k-k_2};k-k_2,\Omega_{k-k_2} )
\]
\[ 
+ \kappa _2 ( k-k_2,\Omega _{k_2}+\Omega_{k-k_2};k_2,\Omega _{k_2} ) ] , 
\]
\[
\omega _k=\Omega _k-sign\left( k\right) \sqrt{\omega ^2-\lambda n}. 
\]

\noindent Averaging of the equation for $A_k$ yields the kinetic equation
for waves

\[
\left( \frac \partial {\partial \theta }+\Omega _g\frac \partial {\partial
\sigma }-2\Gamma _k\right) I_k= 
\]

\[
=2\pi \sum_{k+k_1=k_2+k_3}\left| S\left( k,k_1,k_2,k_3\right)
\right| ^2 
\]

\[
\times \delta \left( \omega _k+\omega _{k_1}-\omega _{k_2}-\omega
_{k_3}\right) 
\]

\begin{equation}
\times \left(
 I_{k_1}I_{k_2}I_{k_3}+I_kI_{k_2}I_{k_3}-I_kI_{k_1}I_{k_3}-I_kI_{k_1}I_{k_2}%
\right) , 
\end{equation}

\noindent where

\[
\left\langle A_kA_{k_1}\right\rangle =I_k\delta \left( k+k_1\right) . 
\]

\noindent Dimensional analysis of the kinetic equation for waves gives the
following fluctuation spectrum law of Kolmogorov type

\begin{equation}
I_k\sim const*k^{-7/3}. 
\end{equation}

\par
We note that this power law spectrum, which in this case is due to a
resonator, would indeed lead to the type of
broad fluctuation spectrum seen in experiments.  However, at this time,
a detailed study of the scaling of the observed spectrum has not been
made.  An extension of the above work to consider bunched beams, and
other types of machine impedance would be very worthwhile.
  
\section{Simulations}

\begin{figure}[t]
\setlength{\figlen}{2.5in}
\setlength{\figwid}{3.0in}
\centerline{\psfig{file=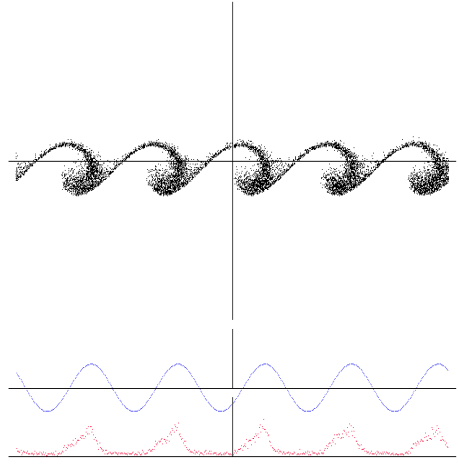,height=\figlen,width=\figwid}}
\caption{Simulation after 500 time steps for the model 
problem listed in Table 1. The upper portion represents
$\varepsilon - \theta$ phase-space.  The middle curve is the cavity voltage as
applied to different portions of the beam and the bottom curve is the
projection of phase-space on the theta-axis, representing the instantaneous 
current.  The resonator wakefields have caused bunching which shows
significant particle trapping and consequent wave overturning.  The
particle current shows strong local intensification of the beam density
corresponding to the solitons.}
\label{figure:sims}
\end{figure}

\begin{figure}[t]
\setlength{\figlen}{2.5in}
\setlength{\figwid}{3.0in}
\centerline{\psfig{file=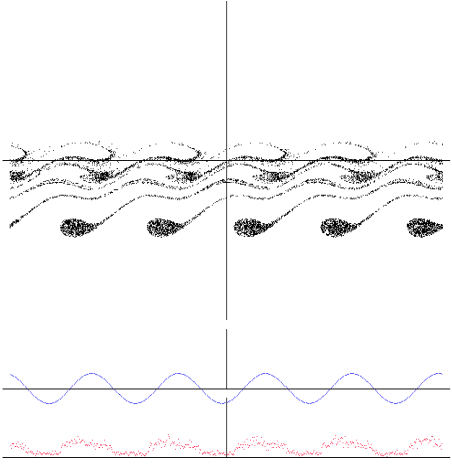,height=\figlen,width=\figwid}}
\caption{Phase-space after 1000 time steps.  Portions of the originally 
trapped particle population have decelerated from the core due to the 
finite resistivity of the wakefields.  The solitons remain, however,
well-organized as they experience deceleration.  Both the cavity voltage
and the density perturbations have decreased, but persist for long times.
The diagrams have the same meaning as in the previous figure.}
\label{figure:sims1}
\end{figure}

\begin{figure}[t]
\setlength{\figlen}{2.5in}
\setlength{\figwid}{3.0in}
\centerline{\psfig{file=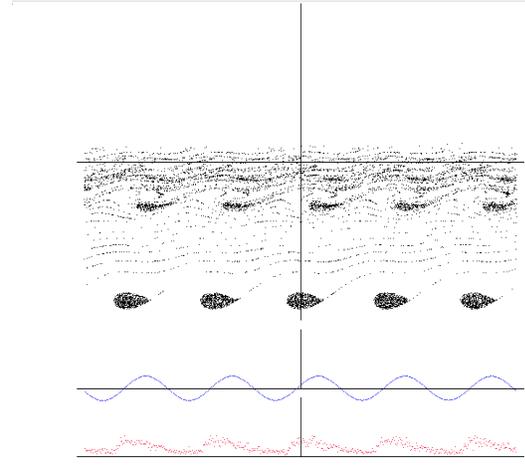,height=\figlen,width=\figwid}}
\caption{Phase-space after 2000 time steps. The beam core is now 
largely decohered, however, the solitons continue to decelerate
slowly and maintain the remaining cavity voltage well off resonance. 
The diagrams have the same meaning as in the previous figure.}
\label{figure:sims2}
\end{figure}

In this work, we are interested in demonstrating examples of the nonlinear
wave behavior we have described analytically.  No attempt is made to closely model
a real device, though this could be done with the building blocks we provide
here.  We shall concentrate on the longitudinal plane, as explained above,
and carry out simulations exculsively on a coasting, (unbunched) beam, for
simplicity.  We adopt the resonator model described above and follow
approximately the procedure adopted in early simulation work \cite{keil}.
The particle evolution equations are given by

\[
\frac{d \epsilon}{d t} = 2\pi e\omega_s V(t)
\]
\[
\frac{d \theta}{d t} = - \eta \epsilon
\]
\begin{equation}
\frac{d V}{d t} = \frac{\omega_r R}{Q} (I-I_1) - \frac{\omega_r}{Q} V
\label{equation:volt}
\end{equation}
\[
\frac{d I_1}{d t} = \frac{\omega_r Q}{R} V
\]

\noindent
where $\epsilon$ is the energy deviation from the synchronous energy
and V is the voltage induced in a resonator with shunt impedance R,
resonant frequency $\omega_r$ and quality factor Q.  I is the instantaneous
current given as the projection of phase-space onto the $\theta$ axis
\[
I = \frac{e \omega_s}{2 \pi} \int_{-\infty}^{\infty} f d\epsilon
\]
\noindent
The energy dispersion $\eta$ is assumed to be a constant and the particle
distribution is advanced each time step according to Eq. \ref{equation:volt}.
Then the current is computed, followed by the voltage on the resonator using
Eq. \ref{equation:volt}. Results for the case of a coasting 
beam are shown in Figure \ref{figure:sims}
- \ref{figure:sims2}.  The simulation parameters are given in 
Table \ref{tab:simtab}.

%
%

In Fig. \ref{figure:sims}, the phase-space distribution is
shown, initially assumed to be uniform in $\theta$ and Gaussian in energy.  After
500 time steps, the resonator has developed a sinusoidal voltage from 
the initial noise level (due to the finite number of particles) which has
succeeded in bunching a large fraction of the beam and synchrotron motion
in the resulting potential well is taking place.  The synchrotron motion
can also be thought of as synonymous to the wave-breaking process which 
has been described in plasma physics \cite{dawson}.  As time proceeds, Fig. 
\ref{figure:sims1}, shows
a deceleration of the trapped particles from the core of the beam, and the
decelerated portion remains well-organized, and even intensifies as its length
is foreshortened.  The voltage in the resonator then becomes phase-locked 
to the 'droplets' and the voltage amplitude oscillates as they move in and
out of phase with the remaining coherent structure in the beam's core.  
We note that the droplets thus formed bear the characteristics of the solitons
discussed in the previous section, remaining self-organized for long times.
\par
\begin{figure}[t]
\setlength{\figlen}{2.5in}
\setlength{\figwid}{3.0in}
\centerline{\psfig{file=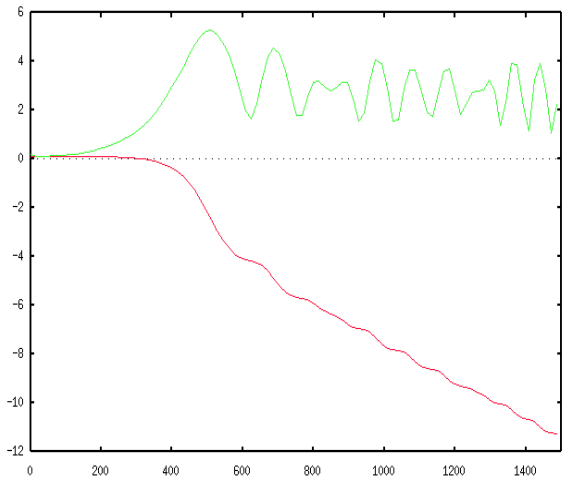,height=\figlen,width=\figwid}}
\caption{Cavity voltage envelope and mean energy deviation.  The
upper curve is the envelope of the cavity voltage, showing the
exchange of energy between the soliton and the wakefield.  The
lower curve is the mean energy of the perturbation, which descends
away from the beam, but eventually reaches an equilibrium (not
shown) where the deceleration ceases.}
\label{figure:envelope}
\end{figure}

As the dissipation in the impedance continues to decelerate the solitons
Fig. \ref{figure:sims2}, the resonator voltage drops due to the high Q value,
or narrow bandwidth, assumed.  This, in turn, reduces the deceleration
rate and the depth of the potential wells that can sustain the solitons.
As such, a steady state can be reached where the remaining trapped particles 
reach a stable equilibrium outside the beam, in accord with the analytic model
for solitary waves.  The envelope of the cavity voltage and the mean energy
deviation of the solitons are shown in Fig. \ref{figure:envelope}, indicating
deceleration of the trapped particles.  A steady state is eventually
reached, though not shown, where the solitons have moved sufficiently off
resonance that the deceleration ceases.
\par
\begin{figure}[t]
\setlength{\figlen}{2.5in}
\setlength{\figwid}{3.0in}
\centerline{\psfig{file=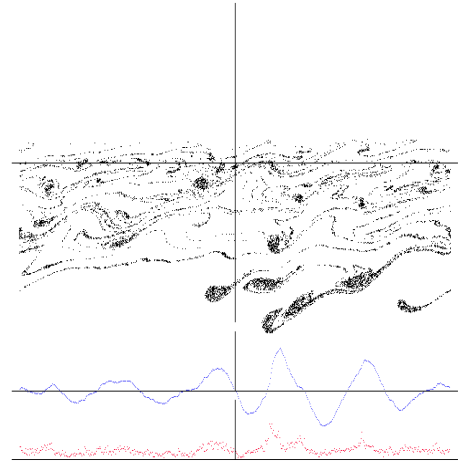,height=\figlen,width=\figwid}}
\caption{Simulation results for low-Q case.  The comparative contribution
of noise to the dynamics is larger owing to the increased bandwidth of
the resonator.  Solitary waves still develop, but the associated 
fluctuation spectrum is broader.}
\label{figure:lowQ}
\end{figure}

We show the final results for a low-Q cavity in Fig. \ref{figure:lowQ}.
These are qualitatively the same as in the previous case, but the structure
of the solitons has taken on a decidedly random character.  This is evidently
due to the fact that the fractional contribution of noise to the cavity
voltage is larger, owing to the wider bandwidth of the cavity, resulting 
in a more random distribution of potential well sizes.  Droplet, or soliton
formation, still occurs, but the resulting fluctuation spectrum is broader.
The onset of the solitary waves can be viewed as a phase transition at the
beam's edge produced by the resonator's wakefield.  This is the case, 
we believe, that is most frequently encountered in actual machines.
\begin{table}[hbt]
\begin{center}
\vspace{2ex}
\caption{Model Simulation Parameters}
\vspace{2ex}
\centering
\centering
\begin {tabular}{|| c | c ||}
\hline
 Parameter & Value \\ 
\hline
 Resonator Impedance & 100 Ohms \\
\hline
 Slip Factor $\eta$  & .001 \\
\hline
Resonator Quality Factor   & 10.\\
\hline
 Typical Beam Energy Spread $\sigma_{\varepsilon}/E_{0}$& $1-4\times 10^{-3}$ \\
\hline
 Number of Particles  & 10000 \\
\hline
\end{tabular}
\label{tab:simtab}
\end{center}
\end{table}
%

\par

\noindent
We note that the fluctuation spectrum associated with the above distribution
is due to the ensemble of strongly nonlinear waves and is likely beyond
the realm of the three-wave interaction described in the scaling law in the
previous section.  An interesting study to be carried out is the experimental
and theoretical determination of the spectral shape in a machine whose impedance is 
well-known.
 
\section{Conclusions}

In this work we have attempted to outline various levels of nonlinearity
in coherent interactions in high-energy beams.  Besides the general
academic interest of nonlinear dynamics, for which high-energy beams
provide an excellent testing ground, there are at least two areas where
the study of nonlinear waves can find application in accelerator physics.
\par
The first is the study of halo formation in which the nonlinear evolution
of coherent fluctuations can lead to soliton, or droplet, formation,
as described in previous sections.  While there is suggestive experimental
evidence that such states can occur, there has been little detailed study
of this phenomenon, and we assert that there is much to be learned about
machine wakefields through the study of solitary waves and their 
interactions.  Specifically, we have only considered the simplest case,
that of longitudinal waves in an unbunched beam, and there are many other
cases of interest in high-energy accelerators.
\par
The second application is the study of non-equilibrium fluctuations
driven by wakefields.  Nonlinear mode-mode coupling permits a frequency
cascade, both toward lower and higher frequencies via separate processes.
The photon distribution that results is an equilibrium between the
nonlinear interactions producing the cascade and weak dissipative
mechanisms.  These mechanisms are assumed to be related to the broadband
impedance of the machine, though other mechanisms may also be responsible.
We have carried out a model calculation for a specific form of impedance
that yields a specific scaling law for the turbulent spectrum.  A number
of assumptions come into play in the development of this model and the
situation is ripe for careful experimental testing.  The benefit of
this study is the understanding of the significance of low-level turbulence
in the limiting parameters of a given accelerator.

\section{Acknowledgements}

The authors are indebted to Prof. Nigel Goldenfeld for helpful
discussions, and to Dr. David Finley for his enlightened
support of this work.

\end{document}